# Massive production of graphene oxide from expanded graphite

Ling SUN, Bunshi FUGETSU

Graduate School of Environmental Science, Hokkaido University, Sapporo 060-0810, Japan

**Abstract:** In a deviation from the conventional Hummers method, a spontaneous expansion approach was introduced with expanded graphite as the precursors. The intercalating agent ($H_2SO_4$) was able to penetrate into the expanded graphite; this had further expanded the graphite and as a result, a foam-like intermediate was produced. The foam-like graphite was more easily oxidized in reaction with the oxidant ($KMnO_4$) to form graphene oxide (GO). Fully exfoliated GO was obtained with expanded graphite having the median diameter ~ 15 μm as the precursors. This procedure was much safer and productive in scalable applications than the conventional Hummers methods.

**Key words:** Expanded graphite; Graphene oxide; Spontaneous expansion; Massive production

## 0  Introduction

Graphene oxide (GO) is of great interest due to its low cost, easy access, and widespread ability to convert to graphene.[1-5] Scalability is also a much desired feature. At present, a conventionally-modified Hummers method is the primary method for preparing GO. Graphite is commonly chosen as the starting material due to its availability and low cost. Proportional amounts of oxidants, such as potassium permanganate, sodium nitrate, and concentrated sulfuric acid, are mixed in order with the graphite. Subsequently, a three-phase procedure is conducted with low, mid, and high temperature reactions, each occurring separately at scheduled times. The graphite was oxidized to GO through these procedures. A large number of oxygen-containing functional groups have been introduced onto both sides of a single graphite sheet (namely, graphene). The implantation of functional groups overcomes the inter-sheet van der Waals force and enlarges the interlayer spacing. The sheets in such an expanded structure are then easily pulled open using an external force such as sonication. That is, the expanded graphite is exfoliated into multi-layered or even single-layered sheets. Generally, the oxidized graphene sheets, namely, GO, acquire multiple defects and the degree of the defects is subject to the additive amount of oxidant and the oxidizing time. [6, 7]

In regard to the oxidants, some research groups, including that of James M. Tour, excluded sodium nitrate as an additive due to its negligible role in graphite oxidation[8]. A mixture of $H_2SO_4/H_3PO_4$ (9:1 volume ratio) instead of only $H_2SO_4$ resulted in increased hydrophilic and oxidized GO without the emission of toxic gas, which differs from the traditional Hummers method.[8] Of these typical processes, the throughput normally benefits from the preparation with a prolonged stirring over time (over 12 h or even 2 d).[3,8] In addition, proper preparation requires that particular attention be paid to some key steps, for example, the addition of water for high temperature hydrolysis. Unintentionally, the mixed highly explosive oxidants can decompose in an exothermic process or as a large explosion; therefore, off-the-rack methods starting with graphite are still flawed.

As previously reported, there exists a side effect which influences the oxidization, such that graphite crystal at different sizes would affect the exfoliation of their own subject to the force balance between the sheet-implanted functionalities.[9] Difficulties have always been encountered with graphite having the larger lateral sizes and the higher crystallinity.[10] Thus, size-dependent modifications to the conventional preparation are favorable in terms of cost effectiveness.

Expanded graphite has long been commercially available and its sizes have been optionally chosen. In this paper we report on synthesizing GO with increased safety and productivity by applying commercial expanded graphite having particle sizes in a moderate range (the median diameter D50 ~ 15 μm) as a raw material and with a newly introduced method. For this purpose we introduced a spontaneous-expansion-step modification,

---

**Corresponding author**  E-mail: captainsun@ees.hokudai.ac.jp

which resulted in a preparation that strongly contributes to safety, efficiency, and productivity. Comparisons were conducted among three raw materials with different sizes. Meanwhile, characterization was also performed on the self-assembled GO film. This promising method has applications to the scalable industrialization of graphene.

# 1 Experimental

## 1.1 Materials and reagents

Commercial expanded graphite (EC300 with D50 ~ 50 μm, EC1000 with D50~ 15μm as manufactured) was purchased from Ito Kokuen Co., Ltd, Mieken, Japan. Highly oriented pyrolitic graphite (HOPG) was acquired from Bay Carbon, Inc., Michigan, USA. Other chemicals unless specifically noted were from Wako Pure Chemical Industries, Ltd., or Sigma-Aldrich Inc., Japan.

## 1.2 Preparation of GO

Some modifications were made to the Hummers method[11] and applied to the preparation of GO from industrially expanded graphite. In a typical reaction, potassium permanganate (15 g) and expanded graphite (5 g) were initially stirred until homogeneous. Then, in a bottom-round flask (500 mL) with ice-water bath, concentrated sulfuric acid (98 %, 100 mL) was added to the mixture with continuously stirring until a uniform liquid paste was formed. Then the water bath was removed. The stirring continued until a foam-like intermediate spontaneously formed (around 30 min) at room temperature with a large volumetric expansion. Deionized water (400 mL) was added, and rapid stirring was restarted to prevent effervescing. Next, the flask was placed in a 90 ℃ water bath, and after 1 hour a homogeneous suspension was obtained that was dark yellow in color. The suspension was then filtered and was subjected to repeated washing and centrifugation (10000 rpm, 2 h per cycle) to remove impurities. To fully exfoliate the GO sheets, the resulting solution was sonicated using a tabletop ultrasonication cleaner (100 W).

For the GO film, the as-obtained GO at a specified volume was filtered through a polycarbonate membrane filter (pore size 0.4 μm, diameter 47 mm, Toyo Roshi Kaisha, Ltd). Then the filter film was dried in an oven (80 ℃) for several hours. The dried film was processed in a $Na_2S_2O_4$ solution (100 mg mL$^{-1}$) at 70 ℃ for 10 min. The film color rapidly changed to a deep metallic gray. The reduced GO film was washed several times using deionized water and dried again in the oven (80 ℃) prior to further characterization.

## 1.3 Characterization

The samples were characterized using various analytical methods such as atomic force microscopy (AFM, Agilent series 5500 AFM instrument using the tapping mode at a scanning rate of 0.5 Hz), Raman spectroscopy (Raman, Renishaw inVia Raman microscope, with an excitation wavelength at 532 nm), Fourier transform infrared spectroscopy (FTIR, FT/IR-6100 FT-IR Spectrometer, JASCO), scanning electron microscopy (SEM, JSM-6300, JOEL, with acceleration voltage of 30 kV ), X-ray photoelectron spectroscopy (XPS, JPC-9010MC,JOEL, using Mg K$\alpha$, $1 \times 10^{-7}$ Torr), thermogravimetric analysis (TGA, TG/DTA 6200, SII Exstar6000, with a heating rate of 5℃ per minute under a $N_2$ atmosphere) and X-ray diffraction (XRD) using RINT 2000 (Rigaku Denki, Ltd, X-ray $\lambda_{Cu\,k\alpha} = 0.154$ nm) .

## 2  Results and discussion

The Hummers method is generally regarded as a universal method for preparing graphite oxide from various graphite sources. However, the method has drawbacks with respect to throughput and safety. For small graphite crystals with a large specific surface area, such as expanded graphite (D50~15 μm), an unpredictably fast exothermic process could be induced by the addition of water, and as it produces heavy purple smoke, it implies a large loss of oxidants. Consequently, the conversion efficiency is significantly reduced; furthermore, this process is both terrifying and dangerous.

This difficulty, in this study, was overcome by developing a spontaneous expansion approach. Actually, intercalation of graphite is required for the subsequent exfoliation when preparing GO. $H_2SO_4$ is a most common intercalating agent. Under identical conditions, as compared in Fig. 1, graphite with relatively small lateral sizes would help save considerable time because of a less resistance occurred than in the case of large graphite. Further, if graphite is pre-expanded with a larger inter-layer space, then the intercalation would process faster. Meanwhile we noted the intercalation emits heat as well as produces water. With these points of view, expanded graphite which is cheaply commercially available was used as the starting material. And at the same time, given a spontaneous full intercalation and the evaporation of water at room temperature (Fig. 2), an easily-handled graphite-intercalating intermediate was obtained.

In contrast to the conventional preparation, features of the proposed approach included: (1) full mixing of graphite and potassium permanganate in advance to the addition of $H_2SO_4$ to ensure homogeneity and completeness of the subsequent reaction, excluding sodium nitrate; (2) a mid-temperature reaction performed once the above mixture has reached to a uniform paste in an ice bath; (3) the mid-temperature reaction, which was free from homothermal measurements at room ambient, terminated by a spontaneous volumetric expansion so that the stirring was stopped—this is referred to as "spontaneous-expansion-step"; (4) the ratio of sulfuric acid to graphite is reduced to about 20:1 (v/w), which differs from 23:1 reported in other literature[12, 13]; and finally, (5) a high-temperature reaction (hydrolytic action in 90 ℃ water bath) customized to 1 h.

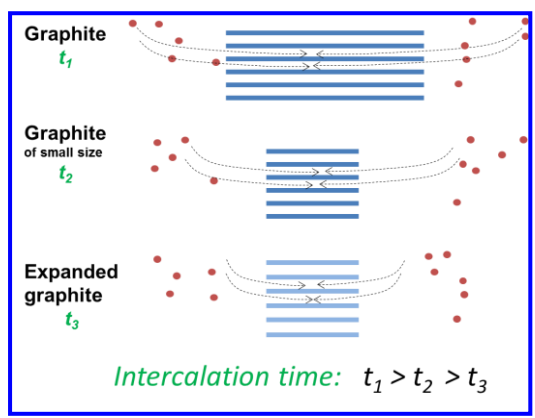

**Fig. 1.**  A schematic of intercalation by $H_2SO_4$ on graphite of different feature.

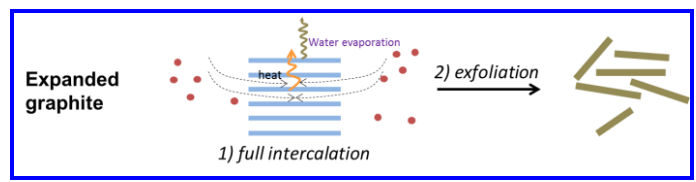

**Fig. 2.** A schematic of the proposed modified method for GO production in large scale from expanded graphite.

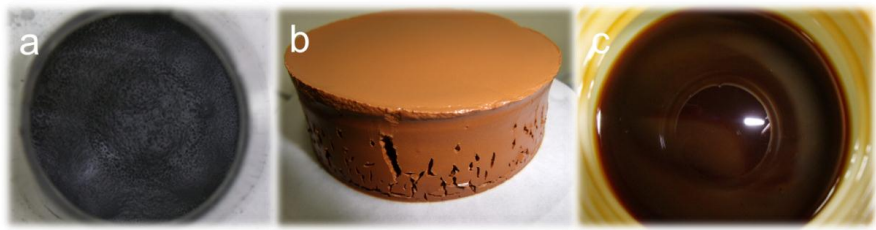

**Fig. 3.** Photos of GO preparation using the proposed method applied on EC1000. (a) The scene at the moment of spontaneous-expansion: a foam-like volumetrically expanded graphite intermediate. (b) The high-temperature hydrolytic reaction product: a homogeneous light-brown GO "cake." (c) Stock purified GO which is dark brown after repeated washing and centrifugation.

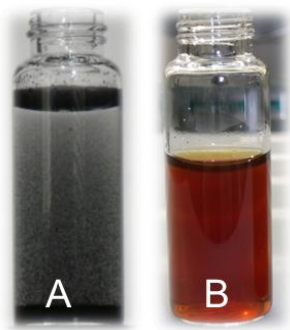

**Fig. 4.** (A) The diluted expanded graphite suspension (EC1000) and (B) Diluted GO solution.

As seen in Fig. 3, the expanded graphite (EC1000) reached the as-mentioned "spontaneous-expansion-stage" soon after its mid-temperature reaction started (less than 30 min). Benefitting from the "foam-like intermediate", water addition was free from security and oxidization failures. Followed by the high-temperature hydrolysis, the "foam-like intermediate" was finally transformed into a suspension with both color and volume distinctly changed. It is important to note that before achieving the "cake" (Fig. 3b), a homogeneous solution with no particles precipitating at the bottom indicated a full conversion of graphite into GO. This procedure was highly repeatable. The removal of impurities rendered the solution dark brown in color (Fig. 3c). A concentrated GO solution was obtained with colloid-like stability over 10 months.

Note that HOPG is highly hydrophobic, while the expanded graphite (EC1000) has some hydrophilicity due to the preliminary oxidization during the production (Fig. 4A). GO produced using our method from EC1000 was well dissolved in water (Fig. 4B), which depended on sheet surface-decorating hydroxyl and carboxyl groups.[14]

Typical AFM images of the resultant GO before and after the sonication treatment are shown in Fig. 5. Without the sonication, GO sheets in a loose worm-like structure were partially observed. Some single pieces of GO were lying around the "worm" as if they were extracted. Self-lamination occurred for the solution containing the worm-like structure during the long-term storage (left inset of Fig. 5). The sonication has split these "worms-like structures" into sheets structures (right part of Fig. 5). The resultant sheets, as shown in Fig. 6, showed the thickness of about 1 nm which is identical to that of the thickness for the single GO sheet (0.7 ~ 1.2 nm)[8, 9, 14, 15, 16].

Two types of the expanded graphite, namely, EC1000 and EC300 were used. As the manufacturer denoted, both EC1000 and EC300 were made of graphite having pre-expansion and sifting. EC1000 was in smaller size (Fig. 7); this results in more reactive sites to attack for the oxidant. On the contrary, EC300 was in the size similar to that of HOPG, thereby having fewer reactive sites. In fact, the as-prepared "GO cakes" with EC300 and HOPG as the precursors contained many coarse particles, (Fig. 8), indicative of an insufficient graphite exfoliation/oxidation.

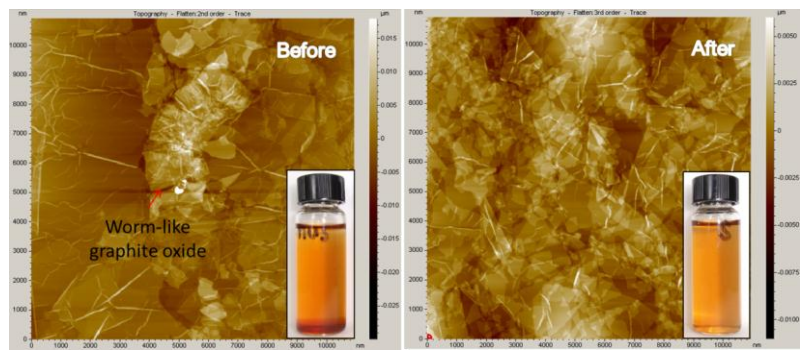

**Fig. 5.** Typical AFM images of GO (left: before sonication, right: after sonication). Insets correspond to their stock solutions after standing over 24 h.

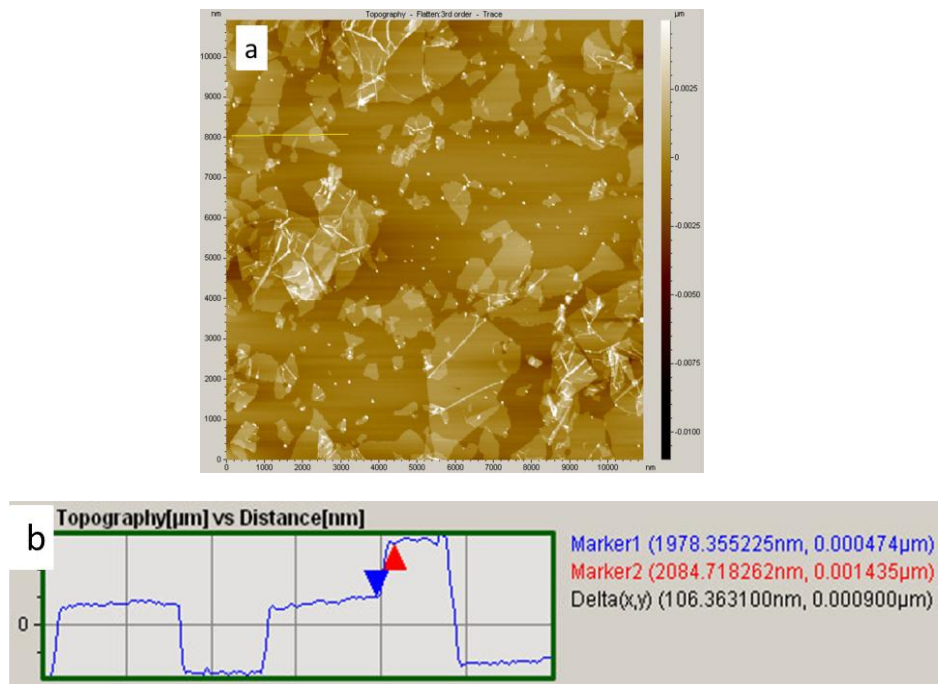

**Fig. 6.** Fully exfoliated GO sheets (a) on freshly mica and a profile image (b) showing the sheet thickness calculated from both trace and retrace modes.

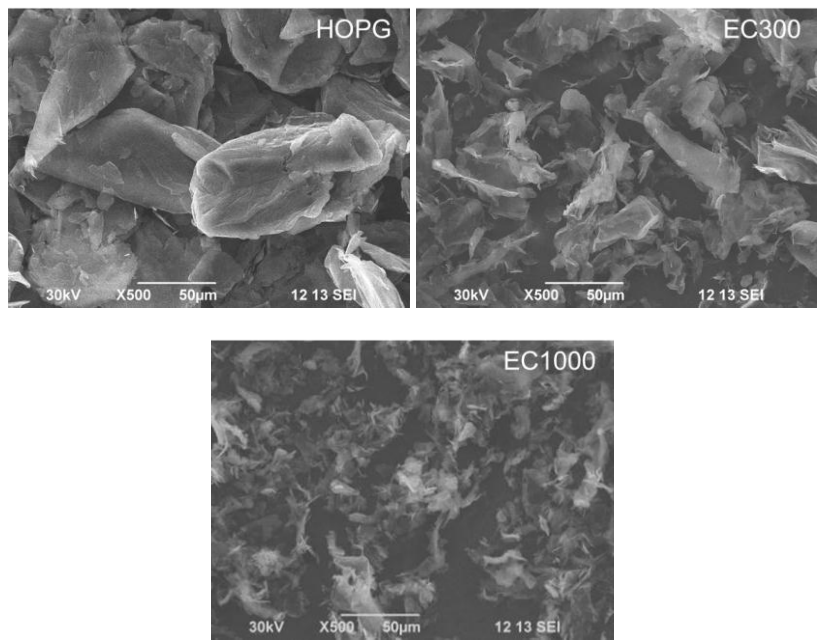

**Fig. 7.** Typical SEM images of graphite sources. HOPG: as-purchased commercially-available highly-oriented pyrolitic graphite (HOPG); EC300 and EC1000: commercially-available expanded graphite of different sizes, D50 ~50 μm for EC300, D50 ~15 μm for EC1000.

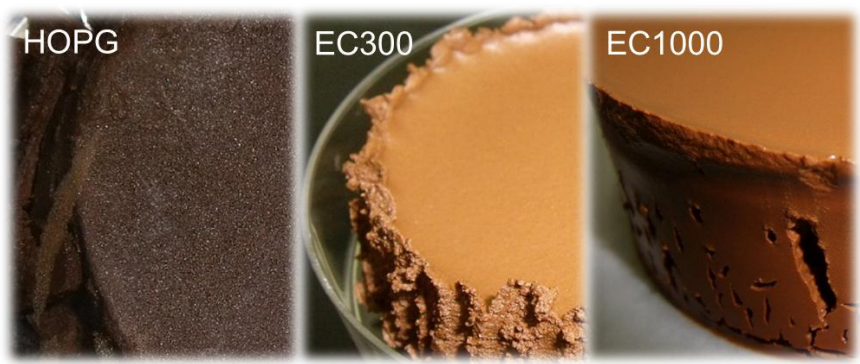

**Fig. 8.** Relevant GO "cake" using the proposed method with HOPG, EC300, and EC1000 as the starting materials.

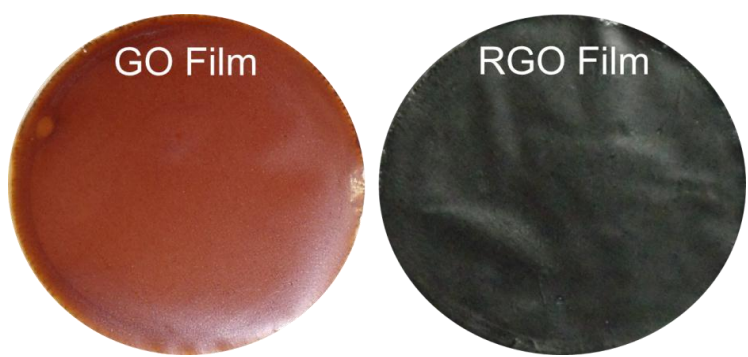

**Fig. 9.** Self-assembled GO film before and after the chemical reduction treatment in a reducing solution containing 100 mg mL$^{-1}$ Na$_2$S$_2$O$_4$ for 10 minutes at 70 ℃.

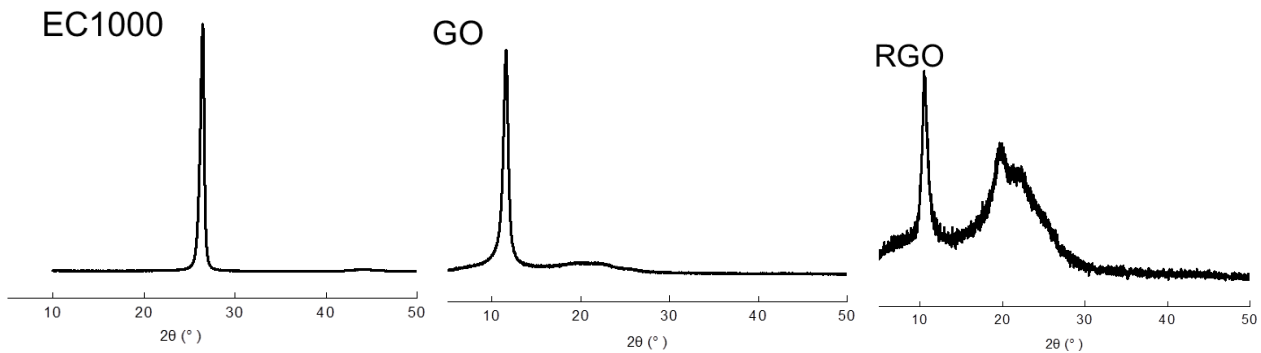

**Fig. 10.** XRD patterns of EC1000 (powder), GO film, and the chemically reduced GO (RGO) film.

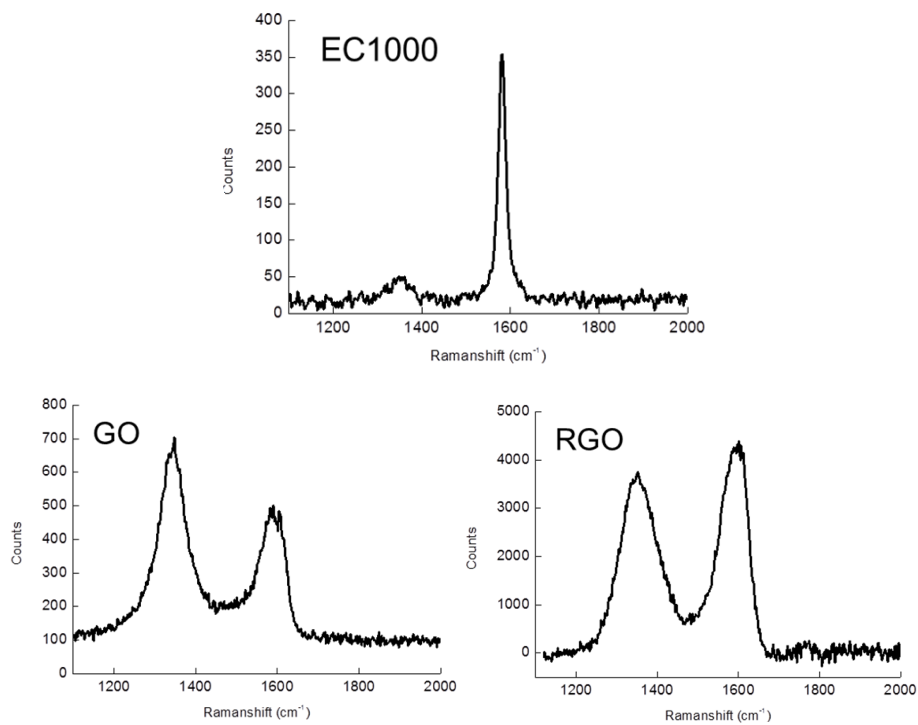

**Fig. 11.** Raman spectra of EC1000 (powder), GO film, and the chemically reduced (RGO) film.

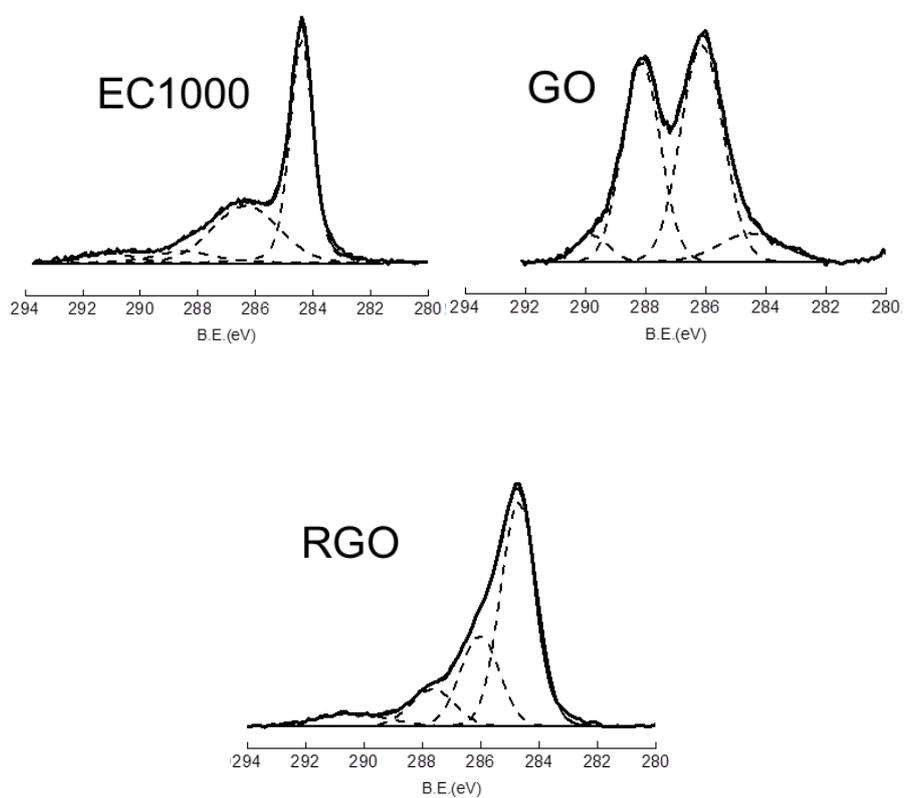

**Fig. 1.** XPS spectra of EC1000 (powder), GO film, and the chemically reduced GO (RGO) film. Dashed lines represent fitting components constituting the respective spectra.

**Table 1.** Proportions of the typical components.

| Components | B.E. (eV) | EC1000 | GO | RGO |
|---|---|---|---|---|
| CC/CH (%) | 284.5 | 49.8 | 9.4 | 57.8 |
| C-O (%) | 286.1 | 33.9 | 47.2 | 24.7 |
| C=O (%) | 288 | 11.7 | 39.3 | 11.3 |
| COO (%) | 290 | - | 4.1 | 6.2 |
| π→π* (%) | 291.5 | 4.6 | - | - |

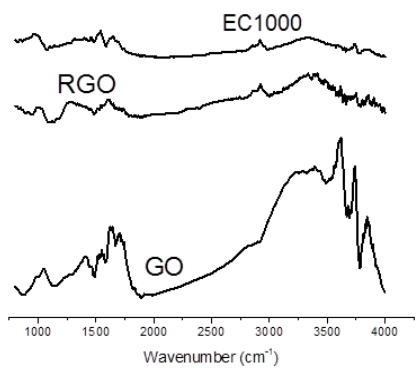

**Fig. 2.** FTIR spectra of EC1000 (powder), GO film, and the chemically reduced GO (RGO) film.

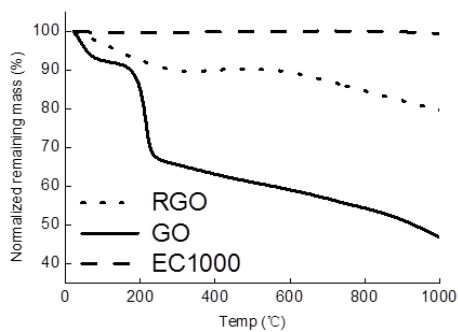

**Fig. 3.** TGA data on EC1000 (powder), GO film, and the chemically reduced GO (RGO) film under nitrogen conditions.

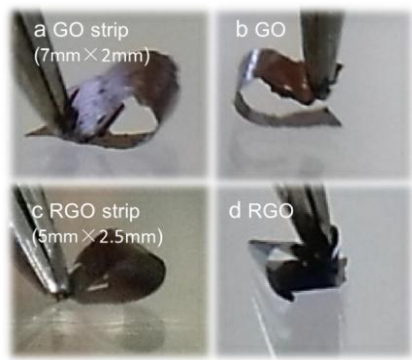

**Fig. 4.** Bending test of the GO film and the chemically reduced GO (RGO) film with confined sizes. The film samples were bent to the left (a, c) and the right (b, d).

Self-assembled GO layered films have received considerable attentions.[16] Developments in this form of GO are ongoing, for example, to make paper-like materials [16-18] for creating supercapacitors[19, 20]. GO prepared in this study was found to be highly applicable for making the GO films; Fig. 9 shows a typical GO film obtained using filtration. The GO film was finally converted in to the graphene film through a chemical reduction. Electric resistivity of the chemically reduced GO film was found at $1.8 \times 10^2$ Ω cm$^{-2}$ level. XRD patterns (Fig. 10) showed the trace of the structural changes. The precursors (EC1000) had a typical peak positioned near 26°, indicating a $d$-spacing of 0.34 nm. The $d$-spacing, however, was 0.76 nm for the GO film. A structural recovery occurred to the chemically reduced GO film was observed. The functional groups had been removed due to their less proportion than GO (see below in XPS analysis) and the van der Waals force-induced inter-sheet re-stacking dominated. The appearance of a broad peak near 22° and the slight right shift of the typical GO peak are assumable due to the incomplete recovery of the original crystal structures.

Characteristic peaks in the Raman spectra varied as the GO underwent the chemical reduction. The first-order scattering of the $E_{2g}$ phonon of $sp^2$ carbon atoms was reflected by a peak around 1581 cm$^{-1}$, historically named the G peak; similarly, a breathing mode of *k*-point photons of $A_{1g}$ symmetry around 1348 cm$^{-1}$ was named the D peak.[21] The ratio of $I_D/I_G$ was designated as the degree of disorder, such as defects, ripples and edges.[21] Fig. 11 shows typical analytical data on Raman spectra for EC1000, the GO film, and the chemically reduced GO film. The $I_D/I_G$ ratios were found to be 0.32, 2.08, and 1.95, respectively.

C1s XPS was performed to acquire the detailed information regarding the functional groups (Fig. 12). The peak at the binding energy (B.E.) around 284.5 eV was assigned to CC/CH; 286.1, 288, and 290 eV were attributed to C-O, C=O, and O=C-O, respectively. Note that for EC1000, the peak at 291.5 eV was assignable to the $\pi \rightarrow \pi^*$ transition of the aromatic C-C bonds. Changes of ratios for these key functional groups for the precursor (EC1000), GO film and the chemically reduced GO film are summarized in Table 1. The overall ratio of the oxygen-containing groups, namely, C-O, C=O, and O=C-O, was found to be 90.6% for our GO film, which was comparable or even higher than that of the ratio for GO films reported in other works[8, 22].

It is noteworthy that the expanded graphite (EC1000) showed a crystallinity as perfect as conventional graphite in the XRD pattern; while with XPS, it appeared to be somewhat oxidized. Thus it could be concluded that the "pre-expanded" graphite was partially oxidized over the outer layer surfaces and/or the edges but the internal plane structures retained unchanged.

The IR spectra of EC1000 and the chemically reduced GO were similar (Fig. 13), specifically in the relatively sharp peak around 3340 cm$^{-1}$, interpreted as the vibration of C-OH. In contrast, for GO, a very broad peak (3200~3400 cm$^{-1}$) was generated from the stretching vibrations of -OH from COOH, C-OH, and H$_2$O; an adjacent sharp peak at 3618 cm$^{-1}$ was also assigned to the vibration of C-OH. In addition, GO exhibited a very strong C=O peak about 1736 cm$^{-1}$, compared with EC1000 and the chemically reduced GO.[21] Fig. 14 shows typical TGA data obtained under nitrogen conditions; for GO, a typical two-step weight loss appeared against the temperature increased, including the liberation of hydrate water and the decomposition of the functional groups. The chemically reduced GO showed much higher stability with less weight loss than the GO, while the EC1000 remained unchanged.

A bending test was conducted to evaluate the flexibility of the GO based films. As shown in Fig. 15, the GO and the chemically reduced GO films could be bent using tweezers at very large angles in both the left and right directions; these films were also very flexible as those shown previously.[23-25]

## 3  Conclusions

We have demonstrated experimentally that starting with suitable resources (expanded graphite, D50 ~ 15 μm), the traditional Hummers method once being modified with a spontaneous expansion process enabled it to become facile and practical for scalable GO production. This method was safe, productive, and most importantly, offers a solid performance. This method introduced in this paper should be applicable for the massive production of GO.